\documentclass[11pt, conference, letterpaper]{IEEEtran}

\usepackage{graphicx,epsfig,amssymb,amsmath,url,latexsym,stfloats,array,bm,bbm}
\usepackage[tight,footnotesize]{subfigure}
\usepackage{makecell}
\usepackage{mathrsfs}
\usepackage{amsthm}
\usepackage{longtable}
\usepackage{amsfonts}
\usepackage{caption}
\usepackage{graphicx}
\usepackage{multirow}
\usepackage{longtable}
\usepackage{afterpage}
\usepackage{subeqnarray}
\usepackage{epstopdf}
\usepackage{empheq}
\usepackage{latexsym}
\usepackage{color, soul}
\usepackage{framed}
\usepackage{lipsum}
\usepackage{color}
\usepackage{stfloats}
\usepackage{setspace}
\definecolor{shadecolor}{rgb}{1,0,0}
\usepackage{cite}
\usepackage{hyperref}
\usepackage{verbatim}
\usepackage{algorithmic}
\usepackage[linesnumbered,ruled]{algorithm2e}
\usepackage[hang]{footmisc}
\usepackage{caption}
\usepackage{cuted}
\usepackage{marvosym}
\IEEEoverridecommandlockouts\IEEEpubid{\makebox[\columnwidth]{\textrm{\Letter} Wen Wu (wuw02@pcl.ac.cn) is the corresponding author of this paper. \hfill} \hspace{\columnsep}\makebox[\columnwidth]{ }}
 \hypersetup{hidelinks}
\begin{document}
\title{
Digital Twin-Assisted Resource Demand Prediction for Multicast Short Video Streaming}
{\setstretch{1.0}
	\author{
	\IEEEauthorblockN{Xinyu Huang\IEEEauthorrefmark{1},  Wen Wu\IEEEauthorrefmark{2}, and Xuemin (Sherman) Shen\IEEEauthorrefmark{1}}
	    \IEEEauthorblockA{\IEEEauthorrefmark{1}Department~of~Electrical~\&~Computer~Engineering,~University~of~Waterloo,~Canada
	    \\\IEEEauthorrefmark{2}Frontier Research Center, Peng~Cheng~Laboratory,~China
	    \\Email: \{x357huan, sshen\}@uwaterloo.ca, wuw02@pcl.ac.cn}
			}
}

\maketitle
\pagestyle{empty}  
\begin{abstract}
In this paper, we propose a digital twin (DT)-assisted resource demand prediction scheme to enhance prediction accuracy for multicast short video streaming. Particularly, we first construct user DTs (UDTs) for collecting real-time user status, including channel condition, location, watching duration, and preference. A reinforcement learning-empowered K-means++ algorithm is developed to cluster users based on the collected user status in UDTs. We then analyze users’ watching duration and preferences in each multicast group to obtain the swiping probability distribution and recommended videos, respectively. The obtained information is utilized to predict radio and computing resource demand of each multicast group. Initial simulation results demonstrate that the proposed scheme can accurately predict resource demand. 
\end{abstract}

\section{Introduction}
With the proliferation of mobile devices and ubiquitous Internet access, an increasing number of individuals rely on short videos to stay up-to-date. According to a recent report from TikTok, the number of active global users per month is expected to reach 1.8 billion by the end of 2023, putting immense pressure on mobile networks~\cite{report}. Multicast technology can effectively enhance the radio resource utilization by utilizing multicast channels to transmit short videos. Due to users’ network dynamics and diversified characteristics \cite{Yu}, short videos usually need to be transcoded into multiple bitrates in the cloud or edge servers to reduce the transmission delay. To facilitate effective multicast transmission and video transcoding, appropriate radio and computing resource reservations are necessary \cite{Zhou}. However, user status, such as channel conditions and swiping behaviors, is relatively dynamic, requiring frequent and accurate multicast group updates. Furthermore, users' swiping behaviors can lead to resource over-provisioning if precached segments are not played. To this end, we need to address the following challenges, i.e., how to accurately and timely cluster users into multicast groups; and how to quantify the effect of users’ swiping behaviors on resource reservation?

The main contributions of this paper are summarized as follows. Firstly, we construct user digital twins (UDTs) to collect user status and propose a two-step method to realize accurate and fast multicast group construction. Secondly, we abstract multicast groups’ swiping probabilities from the watching duration stored in UDTs and utilize them to predict resource demand.

\section{DT-Assisted Resource Demand Prediction Scheme}
\subsection{System Framework}
\begin{figure}[t]
	\centering
	\includegraphics[width=5cm]{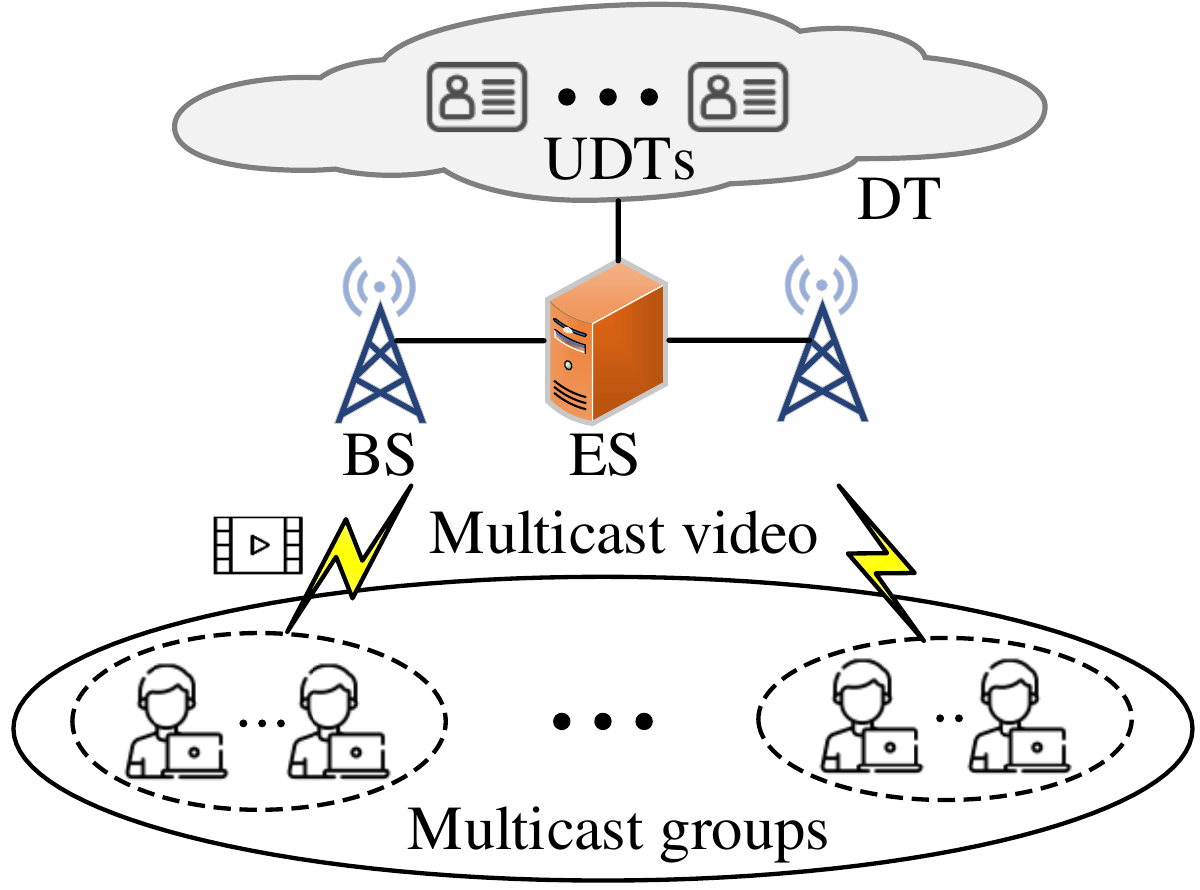}
	\caption{DT-assisted multicast short video framework.}
	\label{fig:framework}
\end{figure}

As shown in Fig.~\ref{fig:framework}, we consider a multicast short video streaming scenario, which consists of multiple base stations (BSs), an edge server (ES), and UDTs.
\begin{itemize}
	
	\item[$\bullet$] BSs: BSs utilize multicast technology to transmit short videos to each multicast group, and collect real-time user status (including channel condition, location, watching duration, and preference) to update the corresponding UDTs' data. Different data attributes are collected with different frequencies.
	
	\item[$\bullet$] ES: The ES connects to BSs and stores popular short videos with the highest representation to reduce frequent content retrieval from remote servers. The stored short videos can be transcoded to a lower representation to adapt to network dynamics.
	
	\item[$\bullet$] UDTs: UDTs are deployed on the edge server to store user status for individual user \cite{shen2021holistic}\cite{globe}.
\end{itemize}

\subsection{Scheme Procedure}
\begin{figure}[t]
	\centering
	\includegraphics[width=8.8cm]{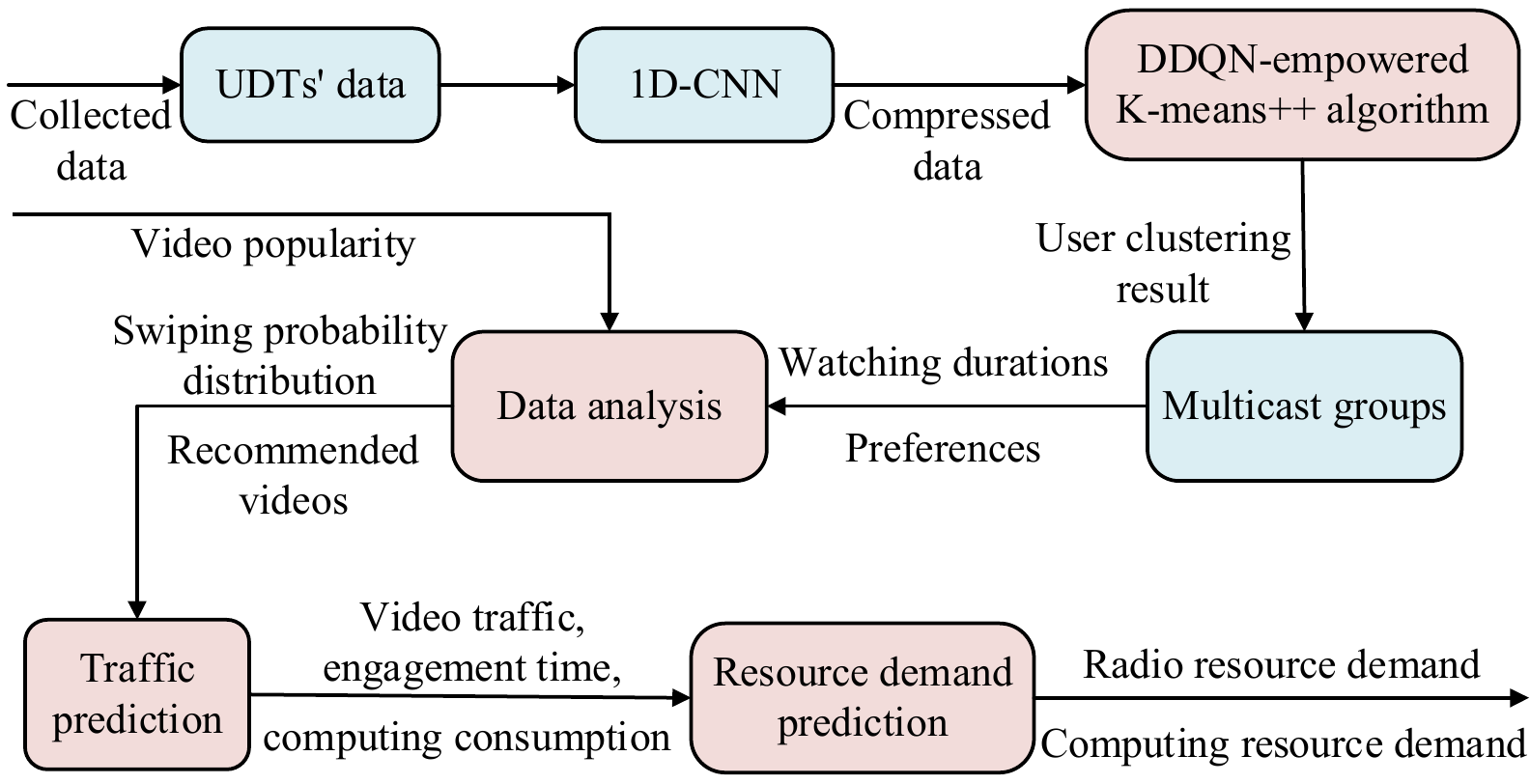}
	\caption{DT-assisted resource demand prediction.}
	\label{fig:workflow}
\end{figure}
To facilitate the accurate resource demand prediction through DT in the considered scenario, the specific procedure includes two main steps, as shown in Fig.~\ref{fig:workflow}.

\subsubsection{Multicast Group Construction} 
Based on the stored user status in UDTs, we analyze user similarity and then divide users with similar status into the same multicast group. The user similarity is defined by the Euclidean distance between any two user status. To this end, we first utilize a one-dimensional convolution neural network (1D-CNN) to compress the time-series UDTs' data. Secondly, we develop a learning-based method for multicast group construction. Specifically, a double deep Q-network (DDQN) is first adopted to determine the grouping number by mining users' similarities. Then, the K-means++ algorithm is utilized to perform fast user clustering based on the determined grouping number.

\subsubsection{Group-Based Resource Demand Prediction} 
Once users are clustered into multicast groups, the corresponding group-level information, i.e., swiping probability distributions and recommended videos, are abstracted. Specifically, users’ watching duration on each kind of video is utilized to update multicast groups’ swiping probability distributions. The recommended videos are updated based on video popularity and users' preferences. Based on the abstracted group-level information, we can analyze multicast groups’ average engagement time, video traffic, and computing consumption to predict radio and computing resource demand.

\section{Initial Simulation Results}
Simulations are conducted to evaluate the performance of the proposed scheme. We adopt the public short-video-streaming-challenge dataset to generate video bitrates and users’ swiping behaviors. The resource reservation interval is 5 minutes. Users’ preferences are updated based on preference labels and engagement time. Users are initially randomly generated in the University of Waterloo campus and then move along different trajectories.
\begin{figure}
	\centering
	\subfigure[]{
		\includegraphics[width=0.23\textwidth]{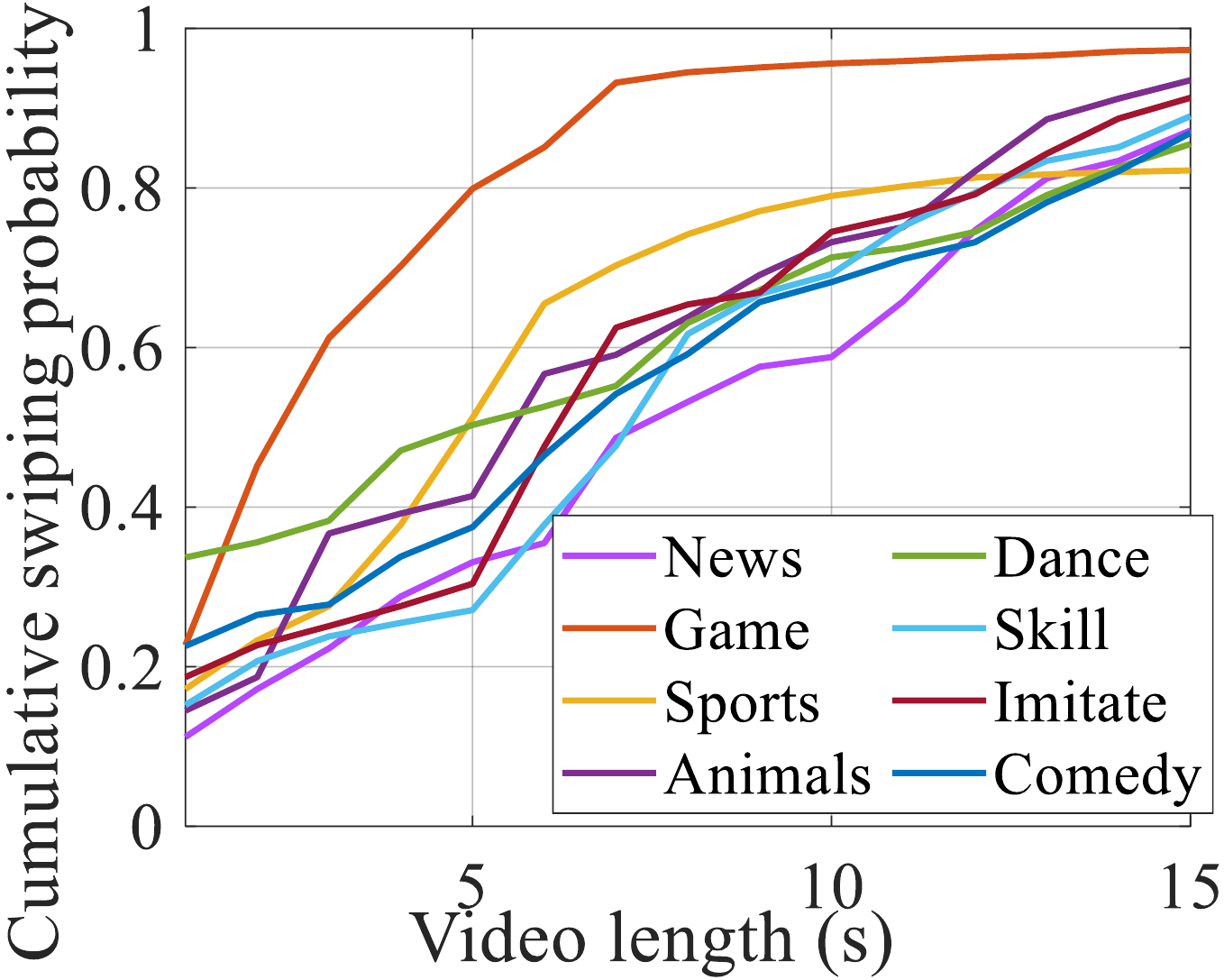}}
	\centering
	\subfigure[]{
		\includegraphics[width=0.23\textwidth]{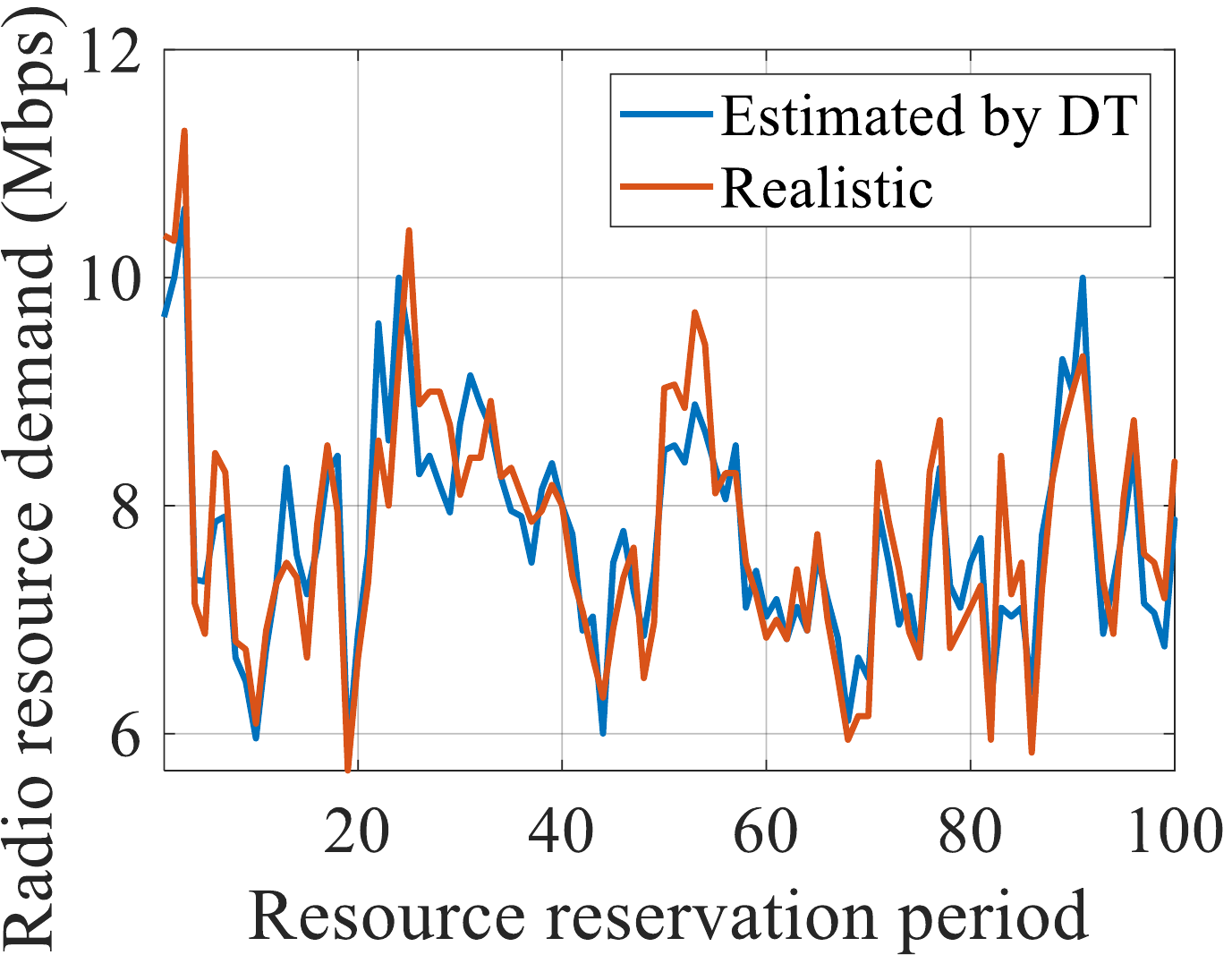}}
	\caption{Swiping probability and radio resource demand.}
	\label{fig:leaving}
\end{figure}

As shown in Fig.~\ref{fig:leaving}, we present the cumulative swiping probability and the radio resource demand for multicast group 1, where users watch News videos most while Game videos least. Based on the abstracted swiping probabilities, our proposed scheme achieves a high prediction accuracy up to 95.04\% on radio resource demand.

\section{Conclusion and Future Work}\label{Conclusion}
In this paper, we have proposed a DT-assisted resource demand prediction scheme in multicast short video streaming, which can effectively abstract multicast groups’ swiping probability distributions and recommend videos. For future work, we will investigate how to effectively reserve radio and computing resources based on the predicted multicast groups' resource demand.

\section*{Acknowledgment}
This work was supported by the Natural Sciences and Engineering Research Council (NSERC) of Canada, and the Peng Cheng Laboratory Major Key Project under Grant PCL2021A09-B2.

\bibliographystyle{IEEEtran}
\bibliography{Ref}
\end{document}